\documentclass[twocolumn,showpacs,prl,nobibnotes,superscriptaddress]{revtex4}
\usepackage{graphicx}

\begin{document}

\title{A novel type of spiral wave with trapped ions}
\author{Yuting Li}
\author{Haihong Li}
\author{Yun Zhu}
\address{School of Science, Beijing University of Posts and
Telecommunications, Beijing, 100876, People's Republic of China}
\author{Mei Zhang}
\address{Physics Department, Beijing Normal University, Beijing, 100875, People's Republic of China}
\author{Junzhong Yang}
\email{jzyang@bupt.edu.cn}
\address{School of Science, Beijing
University of Posts and Telecommunications, Beijing, 100876,
People's Republic of China}

\begin{abstract}
Pattern formation in ultra-cold quantum systems has recently
received a great deal of attention. In this work, we investigate a
two-dimensional model system accounting for the dynamics of
trapped ions. We find a novel spiral wave which is rigidly
rotating but with a peculiar core region in which adjacent ions
oscillate in anti-phase. The formation of this novel spiral wave
is ascribed to the novel excitability reported by Lee and Cross.
The breakup of the novel spiral wave is probed and, especially,
one extraordinary scenario of the disappearance of spiral wave
caused by spontaneous expansion of the anti-phased core is
unveiled.
\end{abstract}


\maketitle

Spiral waves are the most frequently encountered pattern formation
in two-dimensional systems far away from equilibrium. They are
usually thought to be responsible for the patterns in a wide range
of systems (nonlinear optics \cite{leb05}, magnetic films
\cite{sha97}, new chemical systems \cite{agl00}, subcellular
\cite{bre09} biology, and complex plasma \cite{sch11}). The
underlying dynamics supporting spiral waves are usually
oscillatory and excitable. Recently, spiral waves are also
generated in media whose local dynamics is complex periodic or
even chaotic \cite{gor96,par02}. Spiral waves can display a number
of distinct behaviors, some of which are quite complex. The
simplest transition in spiral waves is Hopf bifurcation
\cite{bar92}, which turns a rigidly rotating spiral wave into a
quasi-periodic meandering one.  Being another type of common
transition, the breakup of spiral waves derives from two different
ways: the core breakup and the far field breakup. In the core
breakup, the spiral wave first develops into turbulence near the
spiral core whose mechanism is due to the Doppler effects
\cite{bar93,kar93}. In the far field breakup, the spiral wave
first becomes unstable far away from the spiral core and the
instability underlying it is the absolute Eckhaus one
\cite{ou97,bar99}.

As one type of wave propagation, spiral waves are always explored
in systems where the spatial variable is continuous. However,
spiral waves are also presented in systems of coupled oscillators
where spatial variables are discrete. In the context of coupled
oscillators, the states between adjacent oscillators are not
required to be continuous, which may give rise to some impressing
phenomena on the dynamics of spiral waves. Kuramoto. el. al.
studied a two-dimensional non-locally coupled oscillators and
found the existence of spiral wave chimera \cite{shi04,kur06}. In
a spiral wave chimera, the oscillators in the core region of
spiral wave are desynchronized while those around the periphery of
the core are in synchronization. Martens \emph{et. al.} analyzed
the spiral wave chimera \cite{mar10}. Yang et. al. studied a
two-dimensional locally coupled R\"{o}ssler oscillators
\cite{yang05}. They found a sandwiched spiral wave in which any
two adjacent oscillators are in anti-phase and they attributed the
presence of the sandwiched spiral wave to the shortwave
instability of the homogeneous oscillation of the model
\cite{hea95,hu98}.

Recently, pattern formation in ultra-cold quantum systems which is
of the nature of coupled system has recently received a great deal
of attention. Lee and Cross considered a chain of ions
\cite{lee11}, where dissipation is provided by laser heating and
cooling and nonlinearity comes from trap anharmonicity and beam
shaping. When the nonlinearities and interaction are small
perturbations relative to the harmonic motion of ions, they
derived an amplitude equation for the ions
\begin{eqnarray}\label{eq1}
\frac{dA_{n}}{dt}=ib(A_{n-1}+A_{n+1}-2A_{n})-(1+ic)|A_{n}|^{2}A_{n}+A_{n}.\nonumber\
\end{eqnarray}
$b$ is the coupling between adjacent trapped ions and $c$ denotes
how an ion's amplitude affects its harmonic frequency. The
amplitude equation is similar to the complex Ginzburg-Landau
equation(CGLE). Different from the CGLE which includes both
reactive and dissipative interactions, the above amplitude
equation contains only reactive interaction since the adjacent
ions interact through reactive Coulomb force. Lee and Cross
considered the pattern formation in the system described by
Eq.(1); they found that long-wavelength waves are expected when
$bc>0$ yet very short-wavelength waves for $bc<0$. In the case of
no-flux boundary condition, the only allowed pattern formation for
$bc>0$ is homogeneous oscillation of all ions while very
short-wavelength waves may display a more complicated structure
for $bc<0$. The most surprising discovery by Lee and Cross is
that, when the homogeneous oscillation of ions in the chain for
$bc>0$ is perturbed by a localized pulse of anti-phase
oscillation, the perturbation would probably travel across the
system for a long time before it dies off. This sort of
excitability in an oscillatory medium is put down to the reactive
interaction presented in the system. The new type of excitation of
anti-phase perturbation to the homogeneous oscillation in
Eq.(\ref{eq1}) can also be observed when the system is generalized
to a two-dimensional lattice.

Consider that, for a spiral wave in a two-dimensional coupled
system, its phase singularity (or the rotation center) may be off
lattice and adjacent oscillators on the opposite sides of the
phase singularity are always in anti-phase, which indicates that
there exists a persistent source for anti-phase perturbation. Then
it will be of particular interest to delve how the new type of
excitability reported by Lee and Cross influences the dynamics of
spiral waves on a two-dimensional lattice. In this work, we will
take this quest.

We use a square lattice where ions are trapped to nodes on it. The
system is described as
\begin{eqnarray}\label{eq2}
\frac{dA_{i,j}}{dt}&=&ib(-4A_{i,j}+A_{i-1,j}+A_{i+1,j}+A_{i,j-1}+A_{i,j+1})\nonumber\\
        &&-(1+ic)|A_{i,j}|^{2}A_{i,j}+A_{i,j}
\end{eqnarray}
with $i,j=1,...,N$. Open boundary conditions as in Ref.
\cite{lee11} are imposed upon the system.

\begin{figure}
\includegraphics[width=3in]{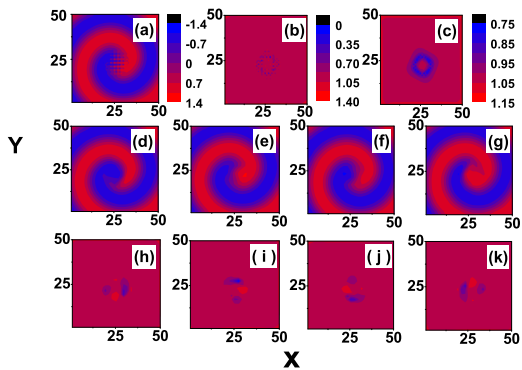}%
\caption{\label{fig_1}(color online) $b=1$ and $c=0.2$. (a) The
wave pattern of SWAPC for $Re(A_{i,j})$. (b) The snapshot of
$|A_{i,j}|$. (c) The snapshot of $\langle|A_{i,j}|\rangle_{t}$.
The middle four panels show the snapshots of $Re(A_{i,j})$ for
different subsets and the bottom four show the snapshots of
$|A_{i,j}|$ for different subsets. (d) and (h) for the subset in
which oscillators locate at positions with $(2i,2j)$. (e) and (i)
The subset with $(2i,2j+1)$. (f) and (j) The subset with
$(2i+1,2j)$. (g) and (k) The subset with $(2i+1,2j+1)$. The color
scale used by (a) [or (b)] are for all plots of $Re(A_{ij})$ (or
$|A_{ij}|$) in this work. $N=50$.}
\end{figure}

To numerically simulate Eq.(\ref{eq2}), we apply the fourth order
Runge-Kutta method with $dt=0.001$. A spiral wave occurs by
initially setting the oscillators at the boundary with
$|A_{i,j}|=1$ and the phases of $A_{i,j}$ running from 0 to $2\pi$
along the boundary. $|A_{i,j}|$ for other oscillators are set to
be  $0$. We first let $b=1$, $c=0.2$ and $N=50$. Stunningly, a
novel type of spiral wave, a rigidly rotating spiral wave with a
core in which adjacent oscillators are in anti-phase, shows up.

\begin{figure}

\includegraphics[width=3.in]{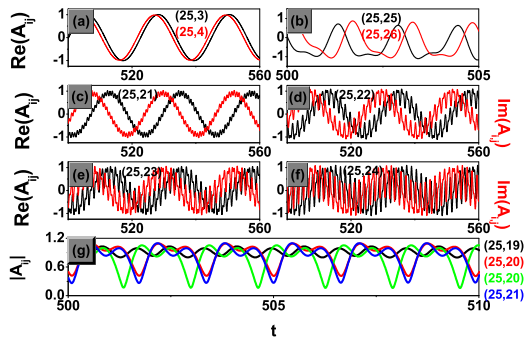}%
\caption{\label{fig_2}(color online) The time evolutions of
different oscillators locating at a line which crosses the core
region of a SWAPC. $b=1$ and $c=0.2$. The subscripts (i, j) in
these plots denote the positions of the oscillators. (a) A pair of
oscillators in the arm region. (b) A pair of oscillators in the
core region. (c)-(f) The successive oscillators in the transition
annulus. The black curves are for $Re(A_{i,j})$ and the red ones
for $Im(A_{i,j})$. (g) The evolutions of $|A_{i,j}|$ for the
oscillators in (c)-(f).}
\end{figure}

For this novel type of spiral wave, two striking features can be
revealed by the snapshots of the wave patterns of the real part of
$A_{i,j}$ ($Re(A_{i,j})$) in Fig. \ref{fig_1}(a) and of the module
of $A(i,j)$ ($|A_{i,j}|$) in Fig. \ref{fig_1}(b). Firstly, the
spiral wave possesses an odd core in which a pattern with the
shortest wavelength appears [See Fig. \ref{fig_1}(a)]; Secondly,
there exists a transition annulus where the oscillation may be
weaker (lower $|A(i,j)|$). It is obvious that the transition
annulus divides the wave pattern into the core region and the arm
region [See Fig. \ref{fig_1}(b)] and, in particular, $|A_{i,j}|$
in the core region is as strong as that in the arm region. In Fig.
\ref{fig_1}(c), we exhibit the pattern for the time average of
$|A_{i,j}|$. Together with Fig. \ref{fig_1}(b), Fig.
\ref{fig_1}(c) illustrates that the novel type of spiral wave is
stable, confirmed by the unchanged location of the core. To be
mentioned, the dimension of the core of the spiral wave may be
defined as that of the transition annulus. Furthermore, we pick up
two pairs of adjacent oscillators: one pair locates in the core
region and the other in the arm region. The time evolutions of
these four oscillators are monitored. As shown in Figs.
\ref{fig_2}(a) and (b), all of them behave periodically.
Nevertheless, it is interesting to find that the phase difference
between adjacent oscillators within the core region is around
$\pi$, which implies the anti-phase characteristic, and is also
responsible for the pattern with the shortest wavelength in the
core region. In contrast, there is only a minor phase shift
between adjacent oscillators within the arm region, which is
induced by the spiral wave propagation. In other words,
oscillators in the arm region perform an in-phased oscillation. In
addition, Figs. \ref{fig_2}(a) and (b) show that the anti-phase
oscillating is greatly faster than that of the homogeneous one,
which is also seen in Ref.\cite{lee11}. Combining the above
observations together, we draw that the novel spiral wave in Fig.
\ref{fig_1}(a) is a rigidly rotating spiral wave with an
anti-phased spiral core (we denote it as SWAPC).

In a CGLE, a rigidly rotating spiral wave always takes the
following form: $A(r,t)=F(r)exp\{i[m\theta+\psi(r)-\omega t]\}$
and $F(r)=0$ at the rotation center (or the phase singularity) of
the spiral wave \cite{ara02}. In other systems such as
reaction-diffusion systems, the similar formulation can be found
provided that the spiral wave is rigidly rotating. Whereas, the
spiral wave presented in Fig. \ref{fig_1}(a) and (b) does not
follow this formulation in the core region. Particularly, the
rotation center of this spiral wave is replaced by an anti-phased
spiral core and the phase singularity of a normal spiral wave is
lost. The statement is supported by the patterns of both
$Re(A_{i,j})$ and $|A_{i,j}|$ for four subsets of the system: the
subset consisting of all oscillators with location
$(2i,2j)(i,j=1,2,......,N/2)$, the subset with $(2i+1,2j)$, the
subset with $(2i,2j+1)$, and the subset with $(2i+1,2j+1)$. As
shown in Fig. \ref{fig_1}(d)-(g), each subset displays a clear
spiral wave pattern which shows that the anti-phased oscillation
of $Re(A_{i,j})$ in the core region is modulated by the in-phased
oscillation in the arm region. Nevertheless, $|A_{i,j}|$ in Fig.
\ref{fig_1}(h)-(k) for each subset substantiates that the phase
singularity for a normal spiral wave is lost. Any subset itself
possess no phase singularity characterized by $|A_{i,j}|=0$ at the
center of the spiral core in spite of the unique singularity a
normal rigidly rotating spiral wave have. For each subset, two
patches with low $|A_{i,j}|$, which situate oppositely to the
center of the spiral core, turn up. It is no other than these
patches in all subsets that contribute to the formation of the
transition annulus.

To further figure out the phase singularity for SWAPC, we regard
how the in-phased oscillation in the arm region passes through the
transition annulus to the anti-phased oscillation in the core
region along a line extending transversely. We present the time
evolutions of successive oscillators on a line down the crossover
area. The results given in Fig. \ref{fig_2}(c)-(f) show distinctly
two periodic components in each oscillator. When the oscillator is
close to the arm region, it is dominated by the in-phased
oscillation which is superimposed with a weak anti-phased
oscillation. As the oscillator approaches the core region, the
component of the anti-phased oscillation grows stronger and
stronger and it becomes predominant. Recalling that the phase
singularity of a normal spiral wave is actually manifested in the
in-phased oscillation, it is the replacement of the in-phased
oscillation by the anti-phased oscillation in the core region that
causes the traditional phase singularity to be lost for SWAPC. To
be mentioned, low $|A_{i,j}|$ ($|A_{i,j}|\simeq 0$) in the
transition annulus is not related to the phase singularity of a
spiral wave. As seen from Fig. \ref{fig_2}(g), low $|A_{i,j}|$ in
the transition annulus just occurs when the anti-phased
oscillation becomes comparable to the in-phased one. Especially,
Fig. \ref{fig_2}(g) shows that $|A_{i,j}|\simeq 0$ in the
transition annulus emerges with an equivalent period as that of
the anti-phased oscillation, which states that the existence of
low $|A_{i,j}|$ in the transition annulus roots in the anti-phased
oscillation and is foreign to the phase singularity. In short, the
phase singularity in a normal spiral wave vanishes for SWAPC and,
instead, an anti-phased core region plays the role of a rotor to
support the spiral wave propagation.

\begin{figure}
\includegraphics[width=3.in]{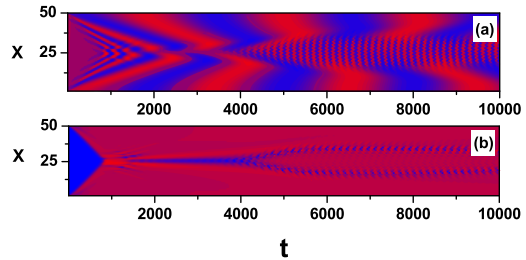}%
\caption{\label{fig_3}(color online) The spatial-temporal plots of
$Re(A_{i,j})$ in (a) and $|A_{i,j}|$ in (b) for the development of
SWAPC. $b=1$ and $c=0.2$.}
\end{figure}

Emphatically, though the wave pattern in Fig. \ref{fig_1}(a) looks
like the spiral wave chimera, they are essentially different.
Firstly, the phases of adjacent oscillators in the core region in
a spiral wave chimera state are unrelated while the phase
difference between adjacent oscillators in SWAPC is kept around
$\pi$. Secondly, as discussed in Ref.\cite{mar10}, the form
$A(r,t)=F(r)exp\{i[m\theta+\psi(r)-\omega t]\}$ is recovered for a
spiral wave chimera provided that $A(r,t)$ is replaced by an order
parameter $R(r,t)$. That is, there exists a well defined phase
singularity in the spiral wave chimera. However, it is quite
different for SWAPC since the ordinary phase singularity has been
substituted by an anti-phased core region where oscillation
amplitude is as strong as that in the arm region.

To get more insight into how the excitability of anti-phased
perturbation to a homogeneous oscillation leads to the formation
of SWAPC, we monitor the evolution of $Re(A_{i,j})$ and
$|A_{i,j}|$ of a chain of oscillators crossing the spiral wave
core from the very initial stage when the novel spiral wave begins
to build. Both spatial-temporal plots of $Re(A_{i,j})$ and
$|A_{i,j}|$ in Fig. \ref{fig_3} show a general scenario towards
SWAPC. An ordinary spiral wave is generated originally; Then, near
the phase singularity of the normal spiral wave, an anti-phased
perturbation comes into being; Ultimately, an anti-phased region
with unchanged size appears and the novel spiral wave is
established. The elaboration above explicitly demonstrates that
the excitability of the anti-phased perturbation from the
in-phased oscillation plays crucial roles on the formation of
SWAPC.

\begin{figure}
\includegraphics[width=3in]{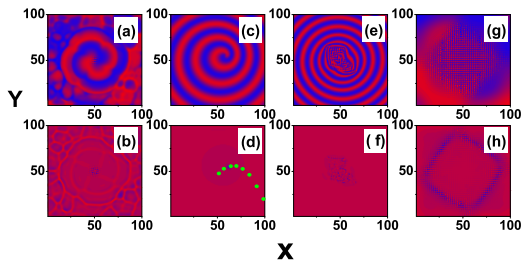}%
\caption{\label{fig_5}(color online) Different types of
instability of SWAPC. The top panels show the snapshots of
$Re(A_{i,j})$ and the bottom $|A_{i,j}|$. (a) and (b) are the
far-away breakup of SWAPC owing to Eckhaus instability. $b=1$ and
$c=-0.33$. (c) and (d) are the evanesce of SWAPC resulting from
spiral wave drifting. The green dots in (d) designates the
trajectory of the successive phase singularities. $b=1$ and
$c=0.8$. (e) and (f) are the novel near-core breakup of SWAPC due
to the spontaneous expansion of the anti-phased core. $b=0.25$ and
$c=0.2$. (g) and (h) are the disappearance of SWAPC on account of
the larger anti-phased core. $b=5$ and $c=0.2$. The size of the
system $N=100$.}
\end{figure}

SWAPC could exist within a large parameter range when $b$ and $c$
alter. The variation of $b$ or $c$ affects on the dynamics of
SWAPC distinctly. Augmented $b$ always leads to the expansion of
the anti-phased core and longer wavelength in the arm region.
Oppositely, with fixed $b$, both the size of anti-phased core and
wavelength in the arm region are independent of the variation of
$c$. In addition, either an outwardly propagating SWAPC or an
inwardly propagating SWAPC wave \cite{van01} can be observed
depending on $b$ and $c$.

To further investigate the dynamics of SWAPC, we investigate how
it breaks up which is always an interesting issue in the field of
pattern formation. We consider a large system with $N=100$. The
results are insensitive to the system size (We have verified this
prediction for different system sizes). In a normal CGLE, the
spiral wave always breaks up through Eckhaus instability. However,
depending on how the parameters vary, SWAPC in the system Eq.
\ref{eq2} may display rich scenes. In this work, merely several
simple situations are taken into account. We prescribe $b=1$ at
first. When $c$ decreases beyond a critical value, the Ekhaus
instability steps in and the normal far-away breakup of spiral
wave appears. A snapshot is shown in Fig. \ref{fig_5}(a) where a
spiral wavelet with an anti-phased core is surrounded by a chaotic
sea. The snapshot of the corresponding $|A(i,j)|$ presented in
Fig. \ref{fig_5}(b) shows that there are many defects outside the
spiral wavelet. As time goes on, the spiral wavelet will shrinks
further and it will die off when the anti-phased core is swallowed
by the defect sea. To be mentioned, the final state after the
disappearance of the spiral wavelet with the anti-phased core is
portrayed by the anti-phase pattern. On the other hand, though
increasing $c$ may also cause SWAPC to vanish, it does not induce
spiral wave breakup. Actually, when $c$ is above a critical value,
SWAPC is replaced by a normal spiral wave in the beginning;
furthermore, to suppress the growth of the anti-phased
perturbation brought about by the phase singularity, the normal
spiral wave has to drift spontaneously. Fig. \ref{fig_5}(c) shows
a snapshot of a drifting spiral wave with normal phase
singularity. The trajectory of the drifting spiral wave is
presented in Fig. \ref{fig_5}(d) where the phase singularities at
successive times are denoted on the plot. Clearly, the drifting
spiral wave dies away in the end due to the collision between its
phase singularity and the boundary. Then, we fix $c$ while change
$b$. For small $b$, one novel instability of SWAPC arises where
the anti-phased core dilates spontaneously and persistently till
overspreads the system and the arm region is eventually drive
away. The snapshots of $Re(A_{i,j})$ and $|A_{i,j}|$ in Fig.
\ref{fig_5}(e) and (f) exhibit a SWAPC with a bulk of anti-phased
core. Different from aforementioned circumstances, there is no
instability of SWAPC is observed by increasing $b$. Nevertheless,
there exist no SWAPC for sufficiently large $b$ for any
finite-size system since the ever swelling anti-phased core.

In summary, we find a novel type of spiral wave when studying the
dynamics of trapped ions on the model system Eq.\ref{eq2}: a
rigidly rotating spiral wave with an anti-phased core (SWAPC). The
formation of SWAPC is mainly due to the excitability discovered by
Lee and Cross. Despite the dynamics of SWAPC as well as its
breakup scenarios has been thoroughly investigated in this work,
there still remain some open problems such as how does the size of
the anti-phased core change and what comprehensive description is
contained in the breakup picture. Furthermore, whether SWAPC
exists in other cold atom quantum systems is also a fascinating
topic.

\end{document}